\documentclass[a4paper]{jpconf}
\usepackage{graphicx}

\usepackage[latin2]{inputenc}
\usepackage{lmodern}
\usepackage[T1]{fontenc} 
\usepackage{pgf}
\usepackage{color}

\newlength{\figurewidth}
\newcommand\const{\rm{const}}

\newcommand\half{\frac{1}{2}}
\newcommand\minihalf{{\scriptstyle{\frac{1}{2}}}}

\newcommand\cm{\mathrm{cm}}
\newcommand\tot{\mathrm{tot}}

\newcommand\Eiav{\bar{E}}
\newcommand\si{\sigma}
\newcommand\scr{\mathrm{scr}}
\newcommand\ii{\mathrm{i}}

\newcommand\ttd{\tilde{t}}
\newcommand\xtd{\tilde{x}}
\newcommand\ptd{\tilde{p}}
\newcommand\Ltd{\tilde{L}}
\newcommand\Htd{\tilde{H}}

\begin{document}
\title{Centre of mass decoherence due to  time dilation: paradoxical frame-dependence}

\author{Lajos Di\'osi}

\address{Wigner Research Centre for Physics, Budapest 114, P.O.Box 49, H-1525  Hungary}

\ead{diosi.lajos@wigner.mta.hu}

\begin{abstract}
The recently proposed centre-of-mass decoherence
of composite objects due to gravitational time-dilation 
[Pikovski {\it et al}, Nat.Phys. 11, 668 (2015)] is confronted
with the principle of equivalence between gravity and observer's
acceleration.  In the laboratory frame, a positional superposition
$\vert x_1\rangle+\vert x_2\rangle$ can quickly decohere whereas in the free-falling frame, 
as I argue,  the superposition can survive for almost arbitrary long times.  
The paradoxical result is explained by the so far unappreciated feature of the proposed model: 
the centre-of-mass canonical subsystem is ambiguous, it is different in the laboratory and  the 
free-falling frames, respectively.

As long as the centre-of-mass motion of the composite object is non-relativistic, 
a simple Galilean-covariant Hamiltonian represents the Pikovski {\it et al} theory with
exactly the same physical predictions. We emphasize the power of this Hamiltonian to
understand essential features of the Pikovski {\it et al} theory and to moderate
a few divergent statements in recent works.
\end{abstract}

\section{Introduction\footnote{
This section, apart from  last paragraph, coincides with \cite{Dio15} rejected by  \emph{Nature Physics}.
}}
Pikovski, Zych, Costa and Brukner (PZCB) have shown that the centre-of-mass of
a composite object  undergoes universal position-decoherence in the 
presence of gravitational field \cite{Piketal15}. 
They proposed an approximate relativistic 
correction of order $1/c^2$ to the non-relativistic many-body Hamiltonian. 
It results in the coupling
\begin{equation}\label{coupling}
-\frac{H_\ii}{mc^2}\left(\frac{p^2}{2m}-mgx\right)
\end{equation}
between the internal Hamiltonian $H_\ii$ and the difference between  
kinetic and gravitational energies of the  centre-of-mass.
Here $m$ is the total mass, $x,p$ are the centre-of-mass canonical 
coordinate and momentum, resp., $g$ is the gravitational acceleration. 
If the object is prepared at rest in superposition 
$\vert x_1\rangle+\vert x_2\rangle$ of two
positions  $x_1,x_2$ at large vertical difference $\Delta x$ then the internal 
degrees of freedom will quickly decohere the said positional superposition.   
The universal mechanism, stemming from time-dilation in Earth
gravity, decoheres composite systems into the
position basis. This happens within standard physics, 
as emphasized by the authors.

I unearth that this decoherence is not independent of the frame of the observer. 
Changing the laboratory frame for the free-falling one, the coupling in question becomes
\begin{equation}
-\frac{H_\ii}{mc^2}\frac{p^2}{2m}
\end{equation}
by the power of the equivalence principle. 
We can (though don't need to) take the same $H_\ii$ in both the laboratory and the free-fall 
frames since its corrections would be relativistic to contain further $1/c^2$-factors. 
If I solve the Schr\"odinger equation in the free-falling frame, the wave function exhibits 
centre-of-mass kinetic energy decoherence, position decoherence is completely absent. 
In the room temperature example of the authors, a gram-scale object prepared 
in the positional superposition $\vert x_1\rangle+\vert x_2\rangle$ at $\Delta x=10^{-6}$ cm 
gets decohered after about a millisecond --- in the laboratory frame. 
It does not get decohered in the free-falling frame as long as the kinetic
energy decoherence remains marginal. Indeed, the dispersion
of $p$ in the initial superposition must be much bigger than  $\hbar/\Delta x$ 
but otherwise it can be kept so small that the kinetic energy decoherence 
would remain ignorable for any conceivable period. Hence the superposition may survive 
practically forever in the free-falling frame. I find it fairly paradoxical. 
Although decoherence is frame-dependent relativistically, 
it  is impossible that a superposition decays in one frame but it never decays in the other.  

The reason of such paradoxical frame-dependence is simple but surprising.
When we change from laboratory to free-fall frame, 
the laboratory momentum transforms like
$p\Rightarrow p+(m+c^{-2}H_\ii)gt+\mathcal{O}(1/c^4)$ --- it drives out from the 
state space of the laboratory centre-of-mass subsystem! In other words, 
the split of the composite system into centre-of-mass and internal canonical subsystems 
is frame-dependent.
The notion of \emph{canonical} centre-of-mass, unlike the notion of
centre-of-mass coordinate, is frame-dependent. Quantum mechanics uses
the canonical notion. Although in the laboratory frame and, alternatively, in the
free-falling frame we are supposed to measure the state of \emph{the} centre-of-mass,
the latter is defined differently in the two frames, 
allowing for the particular inequivalence of decoherence 
calculated in the two frames respectively.
This is a remarkable feature encoded in the theory of PZCB. 

For a related criticism on \cite{Piketal15}, see also ref. \cite{BonOkoSud15}.

\section{Debates, clarifications}
In 2015, PZCB were neither aware nor interested in the
behaviour of their theory in situations different from what they 
considered the important ones exclusively. Indeed, the claim \cite{Dio15} of paradoxical 
frame dependence originated from such applications rather than
from --- as the authors stated  in \cite{Piketal15a} ---  conceptual oversight.
In their concept,  transformation between different frames means
the laboratory and free-falling observers describe the same system
and the same detector. No doubt then, presence or loss of interference fringes  are 
invariant. Just frame dependence of interference fringes originates from another, 
not less sensible and standard concept of frame transformation:
the laboratory observer applies laboratory detector while
the free-falling observer applies free-falling detector. No doubt
again, for a free-falling mass the laboratory detector sees 
PZCB decoherence but the free-falling detector shows perfect 
fringes.  Using different detectors means different experiments, 
as \cite{Piketal15a} pointed it out correctly but that should not have been
a reason to  brush off the surprising outcomes. Recognition and resolution
of the paradox could have been the \emph{productive} alternative. 
For the unsupported (free-falling) mass two major points should have deserved 
attention. 

First, the centre-of-mass reduced dynamics turned out to
be unitary in the free-fall frame and heavily non-unitary in the
laboratory frame. The offered resolution was the following \cite{Dio15}.
The usual invariance of the split 
\begin{equation}\label{split}
\mathcal{H}_\tot=\mathcal{H}_\cm\otimes\mathcal{H}_\ii
\end{equation}
of the total system into centre-of-mass and internal degrees of freedom is lost
due to the coupling (\ref{coupling}) between centre-of-mass and internal degrees of freedom.
Without this coupling $\mathcal{H}_\cm$ is the same in both frames,
$q,p$ and centre-of-mass density matrices transform unitarily from one frame to the other. 
In the PZCB theory, however, $\mathcal{H}_\cm$ becomes different 
for laboratory and free-fall frames, respectively.
Ref. \cite{Piketal15a}, constantly disregarding the  application of the theory
in a different situation from ref.'s \cite{Piketal15}, failed to appreciate the point: 
the frame-dependence of the split (\ref{split}) explains why the \emph{reduced dynamics 
of the centre-of-mass are not unitary equivalent in both frames}. 

Second, it turned out that measuring the same (identically prepared)  object,
a free-falling detector should see perfect fringes while a laboratory detector may
see no fringes at all. That is paradoxical enough unless we consider a
concrete model of detection. This happened finally with the 
lucid proof of Pang, Chen and Khalili \cite{Panetal16}. These authors
considered the superposition of two centre-of-mass wave packets of a free-falling object,
prepared at respective vertical positions $x_1,x_2$ and at common horizontal
momentum $p$. The two wave packets reach the vertical screen through distance $L$ at arrival time $Lm/p$.
If the screen is free-falling together with the mass, the wave packets form the following 
interference pattern on the screen:
\begin{equation}
\const\times\left(1+V\cos\left[\frac{p(x_1-x_2)/L}{\hbar}x_\scr\right]\right),
\end{equation}
where visibility $V$ reaches $1$ in  ideal experiment. (We set $x_\scr=0$ to height $(x_1+x_2)/2)$).
Now let the srceen move at vertical speed $v_\scr$ relative to the mass.
The fringes get shifted on the screen:
 \begin{equation}
\const\times\left(1+V\cos\left[\frac{p(x_1-x_2)/L}{\hbar}\left(x_\scr-v_\scr\frac{Lm}{p}\right)\right]\right).
\end{equation}
This shift does not influence fringe visibility. Now let us insert the relativistic correction $H_\ii/c^2$ of the mass.
The dispersion $\Delta E$ of the internal energy $H_\ii$ means dispersion of arrival times at the screen (i.e:
\emph{dispersion of time dilation} according to ref. \cite{Piketal15})  
resulting in reduction of fringe visibility by the factor
\begin{equation}
\exp\left(-\frac{1}{2}\left(v_\scr\frac{(x_1-x_2)\Delta E}{\hbar c^2}\right)^2\right).
\end{equation}
This decoherence effect is special relativistic, depends on the relative transverse velocity of the mass and the screen,
but unrelated to gravity. If we insert $v_\scr=gt$ corresponding to
the screen supported in the laboratory on Earth, and understand that $t=Lm/p$ is the (mean) time passed bye 
from preparation of the superposition until its detection then we expect the above decoherence factor coincide 
with the prediction of the PZCB theory:
\begin{equation}
\exp\left(-\frac{t^2}{2}\left(\frac{g(x_1-x_2)\Delta E}{\hbar c^2}\right)^2\right).
\end{equation}
In \cite{Piketal17} PZCB  literally distanced their decoherence effects from
those found by ref. \cite{Panetal16} on the moving screen (cf. Fig. 1).  
\begin{figure}[tbp]
\begin{center}
\includegraphics[width=.5\textwidth, angle=0]{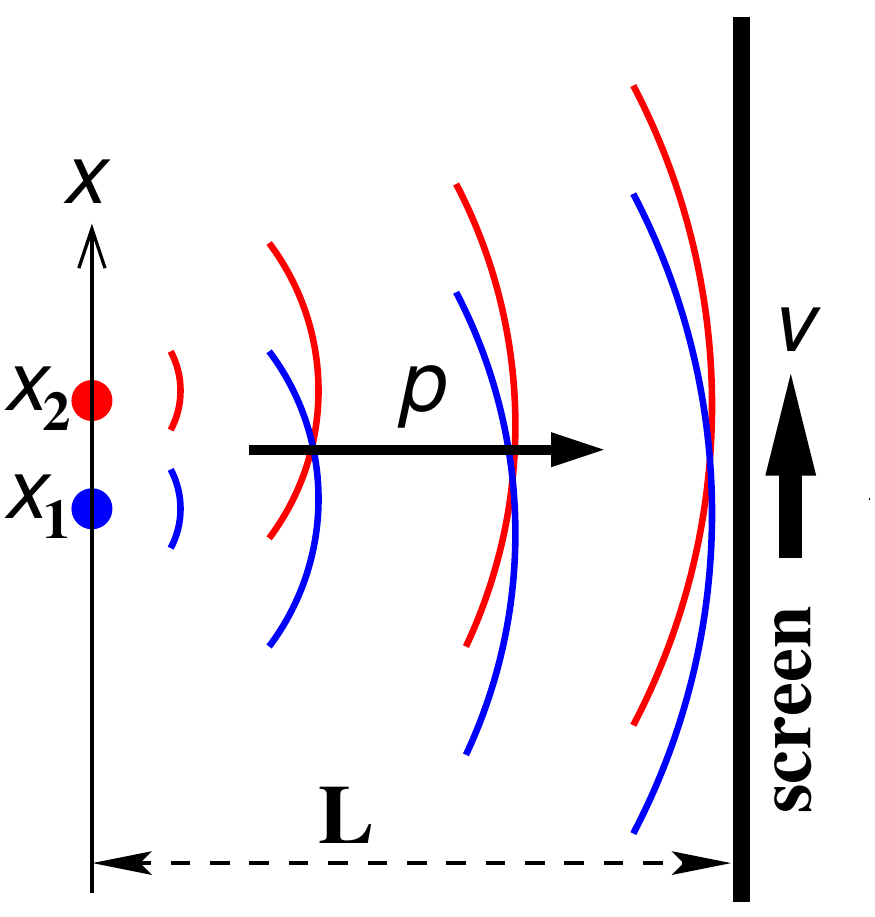}
\end{center}
\caption{Interference of vertical superposition of two centre-of-mass wave packets in inertial motion, 
traveling horizontally at momentum $p$ up to  the screen which is in upward motion at velocity $v_\scr$.  
Vertical shift of the fringes is  $-v_\scr$ times the arrival time.
Since the arrival time $Lm/p$ becomes $L(m+c^{-2}H_\ii)/p$ due to special relativistic time dilation, 
fringe visibility becomes corrupted by the dispersion of the internal energy $H_\ii$. }
\end{figure}

\section{Shortcut to the Pikovski {\it et al} theory}
Let us  start from standard non-relativistic  quantum mechanics of a free composite object:
\begin{equation}
H=\frac{p^2}{2m}+H_\ii.
\end{equation}
The non-standard ansatz is the following: we take the relativistic correction $H_\ii/c^2$
of the mass into the account:
\begin{equation}
H=\frac{p^2}{2(m+c^{-2}H_\ii)}+H_\ii.
\end{equation}
This couples the centre-of-mass and internal degrees of freedom. It is important to emphasize
that we ignore other possible relativistic corrections of the Hamiltonian. The obtained
theory is a non-relativistic theory with a relativistic correction of the mass only.
The Hamiltonian is Galilean-covariant, more precisely, it is covariant under the trivial extension
of the Galilean group \cite{Giu96}. Moreover, it satisfies the Newtonian equivalence principle:
an accelerated frame is equivalent with gravity.
In particular, if we change the inertial (free-falling) reference frame for upward accelerated one
(equivalent to laboratory frame) then, after canonical transformation, the  Hamiltonian reads:
\begin{equation}\label{Hnonrel}
H=\frac{p^2}{2(m+c^{-2}H_\ii)}+(m+c^{-2}H_\ii)gx+H_\ii,
\end{equation}
which, indeed, captures the presence of gravity $g$. The gravitating mass contains the same
relativistic correction that we added to the inertial mass.
The above non-relativistic theory, with its simple Galilean and Newtonian transformation properties,
contains all physics of the PZCB theory as long as the centre-of-mass velocities can be ignored
on the scale of light's velocity $c$, so that we can safely ignore the usual $\propto(p/mc)^2$ 
relativistic corrections of the Hamiltonian.

PZCB \cite{Piketal15} proposed a fruitful relativistic correction to spatial dynamics of 
non-relativistic composite objects. The small parameter of usual relativistic corrections is
$(p/mc)^2$. However, in ref. \cite{Piketal15} there is a second small parameter  $(H_\ii/mc^2)$. \emph{The two 
corrections are independent}. The latter correction, without the usual ones, captures all the novelties
due to relativistic time dilation of the internal `clock' of the composite object.
If the spatial motion is non-relativistic, that has been the case in most, if not all proposals so far,
then the above Galilean-Newtonian Hamiltonian theory is equivalent with the PZCB theory \cite{Piketal15}.

Note that the PZCB coupling (\ref{coupling}) corresponds to the Hamiltonian (\ref{Hnonrel}) expanded
up to terms linear in the small parameter $(H_\ii/mc^2)$. Although the relativistic consistency of higher orders
of  $(H_\ii/mc^2)$ may be questionable, the full Hamiltonian (\ref{Hnonrel}) is convenient for its exact
(extended) Galilean symmetry. 

\section{Epilogue}
The Pikovski {\it et al} theory \cite{Piketal15} is an \emph{original, thoughtful and highly motivating} proposal for the 
effective relativistic coupling between centre-of-mass and internal degrees of freedom, with all its consequences like 
spatial gravitational decoherence, and with all its interpretations in context of 
special and general relativistic time dilation and of quantum foundations.
The PZCB theory had been conceived as an \emph{approximate Lorentz invariant} theory, up to order of $1/c^2$.
This approach has long masked the fact that there is an \emph{exact extended-Galilean-invariant} reduction of PZCB
theory, fully incorporating relativistic time dilation of internal motion and being equivalent to PZCB 
theory {\it as long as the centre-of-mass motion is non-relativistic}. 
Interestingly, in ref. \cite{ZycBru15} Zych and Brukner were fully aware of this theory
but, apparently, they did not accept it as a relevant option.
Recognition would have waived desperate divergences in previous debates, would help ongoing discussions as well. 

Bondar, Okon and Sudarsky \cite{BonOkoSud15} criticized the PZCB theory for it violates Bargman's mass superselection rule \cite{Bar54}. 
The PZCB refutal \cite{Piketal15a} was this: the theory is derived from a
relativistic invariant theory, it is not Galilean-invariant hence not subject to
mass superselection \cite{Bar54}. The precise resolution of the conflict is that mass superposition
is allowed by a trivial extension of the Galilean symmetry, see \cite{Giu96}, under 
which (\ref{Hnonrel}) is exact covariant. This was clear for the authors of \cite{ZycBru15} but, unfortunately, 
the refutal did not mention it.
  
A composite object supported at rest in the laboratory frame, rather than in free-fall,  
was the exclusive object of the original derivation of PZCB centre-of-mass decoherence \cite{Piketal15}.   
A recent work \cite{Toretal17} claimed that the decoherence effect becomes fully 
canceled by the supporting potential. Fortunately, PZCB \cite{Piketal17a} point out correctly
that \cite{Toretal17} constructs the mistaken supporting potential which turns out to be  a 
gravitational field $-g$ just canceling with Earth's gravity $g$. Now, the mistake of \cite{Toretal17}
emerged from the desire to replace the non-relativistic supporting potential well $V(x)$ by
some relativistically covariant supporting field. Had one  started from the Hamiltonian (\ref{Hnonrel}),
an additional supporting potential $V(x)$ would have been added consistently and the 
PZCB decoherence would have been derived correctly and in agreement with \cite{Piketal15}. 

\ack
The author thanks Igor Pikovski, Magdalena Zych, Fabio Costa and {\v C}aslav Brukner
for extended and useful discussions with each of them.  
 This work was supported by EU COST Actions
MP1006, MP1209, CA15220 and the Hungarian Scientific
Research Fund under Grant No. 103917.

\appendix
\section{Derivation of PZCB Hamiltonian from special relativity}
We start with the standard Lagrange function of classical pointlike mass in special relativity,
in Minkowski coordinates:
\begin{equation}
L=-mc\sqrt{c^2-\dot x^2}
\end{equation}
yielding the Hamilton function:
\begin{equation}
H=c\sqrt{m^2c^2+p^2}.
\end{equation}
Consider the same motion in Rindler coordinates $\ttd,\xtd$ (without restriction of generality,
we can take a single spatial dimension). Substituting
\begin{equation}
t=\frac{1}{c}(\xtd+c^2/g)\sinh(g\ttd/c),~~x=(\xtd+c^2/g)\cosh(g\ttd/c)-c^2/g,
\end{equation}
the Lagrange function reads:
\begin{equation}
\Ltd=-mc\sqrt{(c+g\xtd/c)^2-\dot{\xtd}^2},
\end{equation}
yielding the Rindler Hamilton function:  
\begin{equation}
\Htd=c\sqrt{m^2(c+g\xtd/c)^2+\ptd^2}.
\end{equation}
The quantized motion of relativistic point-like masses needs field theory, but the quantum mechanical model 
exists at small velocities $p\ll mc$, with the following Hamiltonian in Minkovski coordinates:
\begin{equation}
H=mc^2+\frac{p^2}{2m}\left(1-\left(\frac{p}{2mc}\right)^2\right),
\end{equation}
and in Rindler coordinates:
\begin{equation}
\Htd=mc^2+\half\left\{\frac{\ptd^2}{2m}+mg\xtd,1-\left(\frac{\ptd}{2mc}\right)^2\right\}.
\end{equation}
If the mass has internal degrees of freedom of Hamiltonian $H_\ii$ then we can introduce 
the relativistic correction $H_\ii/c^2$ to the inertial mass. If we retain terms linear in $H_\ii$,
the above two Hamiltonians take the following forms:
\begin{equation}
H=mc^2+H_\ii+\frac{p^2}{2m}\left(1-\left(\frac{p}{2mc}\right)^2-\left(\frac{H_\ii}{mc^2}\right)^2\right),
\end{equation}
\begin{equation}
\Htd=mc^2+H_\ii+\frac{p^2}{2m}\left(1-\left(\frac{p}{2mc}\right)^2-\left(\frac{H_\ii}{mc^2}\right)^2\right)
                      +\half\left\{mg\xtd,1-\left(\frac{\ptd}{2mc}\right)^2+\left(\frac{H_\ii}{mc^2} \right)^2\right\}.
\end{equation}

\section{Reduced dynamics in laboratory and in free-fall frames\footnote
{As appeared in v2 of \cite{Dio15}.}}
We are going to determine the centre-of-mass reduced density matrix $\rho_\cm(t)$ 
using an uncorrelated initial state:
\begin{equation}
\rho_\cm(t)=\Tr_\ii\left[\e^{-itH/\hbar}\left(\rho_\cm(0)\otimes 
                          \frac{\e^{-H_\ii/k_B T}}{Z}\right)\e^{itH/\hbar}\right].
\end{equation}
Let our Hamiltonian be
\begin{equation}
H=\frac{p^2}{2(m+c^{-2}H_\ii)}+(m+c^{-2}H_\ii)g x+H_\ii
\end{equation}
which yields the coupling used by the authors in the leading order of $1/c^2$. 
We use this non-perturbative form just for convenience of calculations, without
attributing any physical significance to the higher order terms.
[The  Hamiltonian of ref. \cite{Piketal15} must contain a further potential $V(x)$, to
support the two wave packets that are superposed initially.]  
In the free-falling frame we determine $\rho_\cm(t)$, using the Hamiltonian
\begin{equation}
H=\frac{p^2}{2(m+c^{-2}H_\ii)}+H_\ii
\end{equation}
obtained from (B.2) in the free-fall coordinate $x-gt^2/2$.
The laboratory and free-falling frames coincide at $t=0$,
we assume the same initial state, cf. (B.1),  in both frames, respectively. 
It turns out that $\rho_\cm(t)$ may show quick positional decoherence in the laboratory frame 
and practically no decoherence ever in the free-falling frame.

We introduce the following centre-of-mass Hamiltonian in function of the internal 
energy eigenvalues $E$:
\begin{eqnarray}
H(E)&=&\frac{p^2}{2(m+c^{-2}E)}+(m+c^{-2}E)g x\nonumber\\
&\equiv& K(E)+(m+c^{-2}E)g x
\end{eqnarray}
where $K(E)$ stands for the kinetic part.
The reduced state of interest (B.1) can be expressed in  this form:
\begin{equation}
\rho_\cm(t)=\sum_{E}\frac{\e^{-E/k_B T}}{Z}\e^{-itH(E)/\hbar}\rho_\cm(0)\e^{itH(E)/\hbar}.
\end{equation}
The following useful identity holds for the operator part:
\begin{eqnarray}
&~&\langle x_1\vert\e^{-itH(E)/\hbar}\rho_\cm(0)\e^{itH(E)/\hbar}\vert x_2\rangle=\\
&~&~~=e^{-it(m+c^{-2}E)g(x_1-x_2)/\hbar}\times\nonumber\\
&~&~~~~~\times\langle x_1\!\!-\!\!\minihalf gt^2\vert\e^{-itK(E)/\hbar}\rho_\cm(0)
                                                  \e^{itK(E)/\hbar}\vert x_2\!\!-\!\!\minihalf gt^2\rangle.\nonumber
\end{eqnarray}
We can ignore the unitary evolution $\exp[-itK(E)/\hbar]$ if both the coherent momentum 
and momentum uncertainty of the initial state $\rho_\cm(0)$ are suitably small.
This can simply happen to a massive object initially at rest, as we see later.
 So we insert the identity (B.6) with $K(E)=0$ into the expression (B.5) of $\rho_\cm(t)$; we obtain:
\begin{eqnarray}
&~&\langle x_1\vert\rho_\cm(t)\vert x_2\rangle=\nonumber\\
&=&\sum_{E}\frac{\e^{-E/k_B T}}{Z}e^{-it(m+c^{-2}E)g(x_1-x_2)/\hbar}\times\nonumber\\
&~&\times\langle x_1\!\!-\!\!\minihalf gt^2\vert\rho_\cm(0)\vert x_2\!\!-\!\!\minihalf gt^2\rangle.
\end{eqnarray}
We evaluate the pre-factor using the Gaussian approximation of thermodynamic fluctuations. 
The result is an explicit positional decoherence pre-factor:
\begin{eqnarray}
&~&\langle x_1\vert\rho_\cm(t)\vert x_2\rangle=\\
&~&=e^{-\minihalf[c^{-2}\Delta E g(x_1-x_2) t/\hbar]^2 } \times\nonumber\\
&~&~~\times e^{-it(m+c^{-2}\Eiav)g(x_1-x_2)/\hbar}\langle x_1\!\!-\!\!\minihalf gt^2\vert\rho_\cm(0)
                                                                                                     \vert x_2\!\!-\!\!\minihalf gt^2\rangle\nonumber
\end{eqnarray}
where $\Eiav,\Delta E$ are the mean and the fluctuation, respectively,  of the internal energy. 
We read out the decoherence time:
\begin{equation}
\tau_\mathrm{dec}=\frac{\hbar c^2}{\Delta E g\vert x_1-x_2\vert}.
\end{equation}
[This result is equivalent with the one obtained by \cite{Piketal15} 
for the superposition resting in a potential $V(x)$.]
The squared fluctuation is proportional with the heat capacity of the object:
\begin{equation}
(\Delta E)^2=k_B T^2 \frac{d\Eiav}{dT}.
\end{equation}

Let us apply the result (B.8-B.10) to an initial superposition of two Gaussian
wave packets of width $\si$ each, standing at $x_1$ and $x_2$, respectively.
For concreteness, we choose $\si=10^{-8}$ m and $\vert x_1-x_2\vert=10^{-6}$ m, 
guaranteeing the perfect separation of the two
wave packets. [This is a faithful physical representation of the symbolic
superposition $\vert x_1\rangle+\vert x_2\rangle$.] Also we take $m=1$ g and $T=300$ K.
The influence of the kinetic Hamiltonian $K(E)$ is completely ignorable 
as long as $t\ll m\si^2/\hbar\sim 10^{10}$ s. 
Hence, for any conceivable period, the results (B.8-B.10) are correct. 

Guessing the heat capacity of the
$1$ g object by $(d\Eiav/dT)\sim 1$ J/K, we get $\Delta E\sim10^{-4}-10^{-5}$ erg
from eq. (B.10). Therefore eq. (B.9) yields $\tau_\mathrm{dec}\sim0.1-0.01$ ms. 
[If one calculates $\Delta E$ from the microscopic model of ref. \cite{Piketal15}, 
eq. (B.9) recovers its decoherence time $\sim 1$ ms.]

Let us change the reference frame for the free-falling one. 
Using the Hamiltonian (B.3), the counterpart of eq. (B.7) reads:
\begin{equation}
\rho_\cm(t)=\sum_{E}\frac{\e^{-E/k_B T}}{Z}\e^{-itK(E)/\hbar}\rho_\cm(0)\e^{itK(E)/\hbar},
\end{equation}
where $K(E)=\minihalf p^2/(m+c^{-2}E)$ and $p$ is the momentum operator
in the free-fall frame.
We showed above that the chosen initial superposition with the ``wide''
standing wave packets remain practically static against the kinetic Hamiltonian
 $K(E)$. Accordingly, we can set $K(E)=0$ for them and we end up with the
 trivial result:
 \begin{equation}
\rho_\cm(t)=\rho_\cm(0).
\end{equation}
The said superposition of the $1$g object remains static practically forever 
in the free-fall reference frame.

\end{document}